\documentclass[a4paper,fleqn,usenatbib,useAMS,usedcolumn]{mnras}

\usepackage[T1]{fontenc}
\usepackage{ae,aecompl}

\usepackage{graphicx}	
\usepackage[fleqn]{amsmath}	
\usepackage{amssymb}	
\usepackage{gensymb}
\usepackage{hyperref}
\usepackage[nolist]{acronym}
\usepackage{ulem}

\usepackage{etoolbox}
\makeatletter
\patchcmd\@combinedblfloats{\box\@outputbox}{\unvbox\@outputbox}{}{%
  \errmessage{\noexpand\@combinedblfloats could not be patched}%
}%
\makeatother

\acrodef{GW}{gravitational-wave}
\acrodef{NS}{neutron star}
\acrodef{BH}{black hole}
\acrodef{aLIGO}{Advanced LIGO}
\acrodef{AdV}{Advanced Virgo}
\acrodef{PE}{parameter estimation}
\acrodef{IMF}{initial mass function}
\acrodef{MCMC}{Markov-chain Monte Carlo}
\acrodef{COMPAS}{Compact Object Mergers: Population Astrophysics and Statistics}
\acrodef{PDF}{probability density function}

\title[Clustering binary mergers]{Model-independent inference on compact-binary observations}

\author[I.~Mandel et al.]{\parbox{\textwidth}{
Ilya Mandel$^{1}$\thanks{E-mail: imandel@star.sr.bham.ac.uk },
Will M. Farr$^{1}$, 
Andrea Colonna$^{2}$, 
Simon Stevenson$^{1}$, 
Peter Ti\v{n}o$^{2}$, and
John Veitch$^{1}$
}
\vspace{0.2cm}\\
\parbox{\textwidth}{$^{1}$ School of Physics and Astronomy, University of Birmingham, Edgbaston, Birmingham B15 2TT, United Kingdom}\\
\parbox{\textwidth}{$^{2}$ School of Computer Science, University of Birmingham, Edgbaston, Birmingham B15 2TT, United Kingdom}\\
}

\date{\today}

\pubyear{2016}

\begin{document}
\label{firstpage}
\pagerange{\pageref{firstpage}--\pageref{lastpage}}
\maketitle

\begin{abstract}
The recent advanced LIGO detections of gravitational waves from merging binary black holes enhance the prospect of exploring binary evolution via gravitational-wave observations of a population of compact-object binaries.  In the face of uncertainty about binary formation models, model-independent inference provides an appealing alternative to comparisons between observed and modelled populations.  We describe a procedure for clustering in the multi-dimensional parameter space of observations that are subject to significant measurement errors.  We apply this procedure to a mock data set of population-synthesis predictions for the masses of merging compact binaries convolved with realistic measurement uncertainties, and demonstrate that we can accurately distinguish subpopulations of binary neutron stars, binary black holes, and mixed neutron star -- black hole binaries with tens of observations.
\end{abstract}




\section{Introduction}

The advanced LIGO detectors \citep{AdvLIGO} observed the first gravitational waves from a merger of two \acp{BH}, GW150914, on 14 September, 2015 \citep{GW150914}.  This discovery was followed in a few months by another BH-BH merger detection, GW151226 \citep{GW151226}, and a further likely BH-BH candidate, LVT151012 \citep{BBH:O1}.  The BH-BH merger rate inferred from these events implies that tens to hundreds of detections are likely over the next few years \citep{GW150914, BBH:O1}.  Meanwhile, both massive binary evolution models and observations of Galactic binary pulsars and short gamma ray bursts suggest that gravitational-wave detections of mergers of two \acp{NS} and mergers of mixed NS-BH binaries are also likely in the coming years \citep[see][for a review]{ratesdoc}.

Multiple observations should make it possible to address the inverse problem of gravitational-wave astrophysics: to study the currently uncertain massive stellar binary evolution through its evolution end  products --- the population of merging compact remnants.  One approach to this problem involves creating forward models of binary evolution, e.g., via population synthesis Monte Carlo simulations \citep[see][for a review]{PostnovYungelson:2014}, and comparing them to the observed population to constrain the input assumptions, such as the common-envelope physics \citep{Ivanova:2013}.  This approach has been advocated by \citet{BulikBelczynski:2003,MandelOShaughnessy:2010, OShaughnessy:2013,Stevenson:2015} and others.  

While this approach is very promising, existing binary evolution models may not correctly encapsulate the full range of physical uncertainties \citep[e.g.,][]{Dominik:2012,Mennekens:2014,Belczynski:2016,EldridgeStanway:2016,Lipunov:2016}.  Moreover, some of the merging compact binaries could form through channels other than isolated binary evolution via the common-envelope phase, including chemically homogeneous evolution in very close binaries \citep{MandeldeMink:2016,Marchant:2016,deMinkMandel:2016}, dynamical formation in globular clusters, young stellar clusters, or galactic nuclei \citep{Rodriguez:2016,Mapelli:2016,Bartos:2016,Stone:2016}, mergers of population III remnants \citep{Inayoshi:2016} or even primordial black hole mergers \citep{Bird:2016}.  In the possible presence of both systematic model uncertainty and confusion from different formation channels, a model-independent approach to learning from the observed population is desirable.

\citet{Mandel:2015} proposed that clustering on the parameters of observed merging compact-object binaries could provide useful model-independent information about the population.  This clustering is greatly complicated by the limited accuracy with which the masses and spins of merging binaries can be inferred from gravitational-wave observations \citep[e.g.,][]{Veitch:2014,Littenberg:2015,GW150914:PE}.  Nevertheless, \citet{Mandel:2015} suggested that for astrophysically plausible binary populations and realistic measurement uncertainties, a few tens to a few hundred detections should be sufficient to cluster merging binaries into NS-NS, BH-BH, and NS-BH subpopulations, estimating their relative rates to within Poisson uncertainty.  

This paper describes specific algorithms for clustering on the observed merging compact binary population in the presence of significant measurement uncertainty.  We show how the clustering could proceed in practice when subpopulations with distinct mass parameters are brought into contact once the underlying mass distributions are convolved with measurement errors.  We demonstrate the accuracy of the analytical predictions of \citet{Mandel:2015} with a quantitative study.   Our approach  can be trivially extended to include other parameters such as spin magnitudes and spin tilt angles.  

\section{Binary population}

We analyse a realistic population of compact object binaries produced with a population synthesis code that evolves binaries from zero-age main sequence stars through stellar evolution, mass transfer including a possible common-envelope phase, wind-driven mass loss, supernovae, and eventual gravitational-wave driven merger.   For ease of comparison, we use the same simulated binary data set as in \citet{Mandel:2015}. This data set was constructed with the \texttt{StarTrack} code \citep{Belczynski:2008}, using the `Standard' model B of \citet{Dominik:2012}, including the rapid supernova engine \citep{Belczynski:2012,Fryer:2012}, down-selected to binaries potentially detectable by the advanced-detector network as estimated by \citet{Dominik:2014}.  A number of parameters  governing binary evolution are highly uncertain, including wind-driven and luminous blue variable mass loss rates \citep[e.g.,][]{Vink:2001,Mennekens:2014}, mass transfer efficiency \citep[e.g.,][]{deMink:2007SMC}, common-envelope  physics \citep[e.g.,][]{Ivanova:2013}, and black hole natal kicks \citep[e.g.,][]{RepettoNelemans:2015,Mandel:2015kicks}.  Therefore, this model should be viewed only as a realistic illustration for the model-independent inference technique.  

\begin{figure}
\centering
\includegraphics[width=0.45\textwidth]{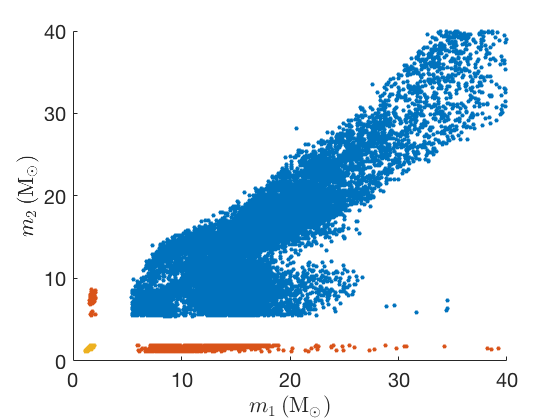}
\caption{Masses of merging compact-object binaries as simulated via population synthesis.  The lower left corner is occupied by NS-NS merging binaries (yellow), the upper right by the more massive BH-BH systems (blue) while the NS-BH population (orange) is asymmetric.}
\label{fig:binaries}
\end{figure}

The modelled population is plotted in \autoref{fig:binaries}.   This population shows clear evidence of a mass gap between neutron stars, whose masses go up to $\sim 2$ solar masses, and black holes, whose masses start at $\sim 5$ solar masses.  This mass gap is a feature of the rapid supernova engine, and reproduces the observed mass gap in neutron star and black hole masses \citep{Ozel:2010,Farr:2010} \citep[but see][]{Kreidberg:2012}.  

\begin{figure}
\centering
\includegraphics[width=0.45\textwidth]{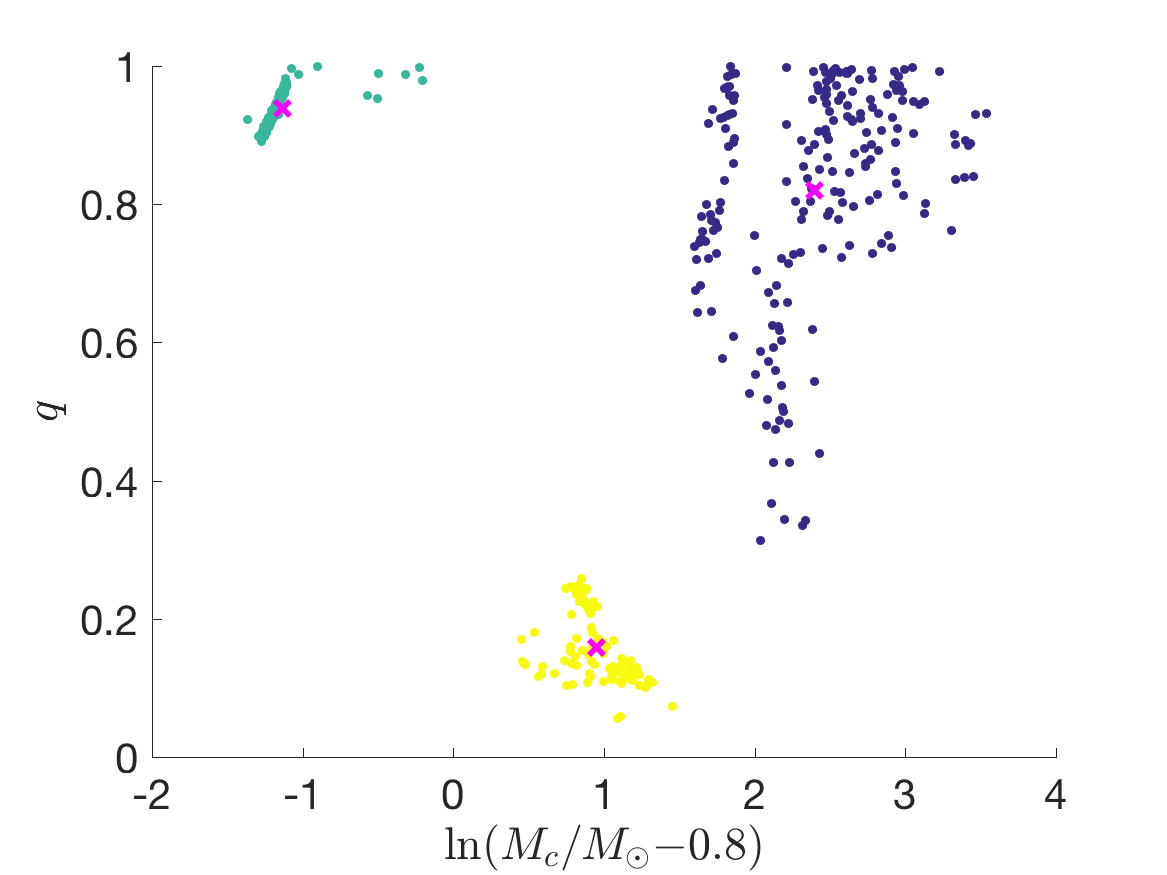}
\caption{K-means clustering on the true masses of 400 simulated compact-object binaries.  The three clusters perfectly match the actual source subpopulations, with the cluster means shown by magenta Xs.}
\label{fig:truecluster}
\end{figure}

In the absence of measurement errors, clustering on a subset of observations should be straightforward, and we demonstrate the feasibility of such clustering in \autoref{fig:truecluster}.  Here, we have chosen 400 merging compact binaries from the population of \autoref{fig:binaries}.  The binaries were randomly drawn with a draw probability of NS-NS, NS-BH, and BH-BH binaries set to 25\%, 25\%, and 50\%, respectively\footnote{These fractions represented an ad hoc choice, not based on the population synthesis model.}; the actual population of 400 selected systems has 23\%, 26\%, and 52\% of binaries of  the three respective types.  These should be interpreted as fractional rates in the observed population, as we do not model selection effects here \citep[cf.~][]{Mandel:2016select}.  Our goal is to extract these arbitrarily chosen relative contributions of the three subpopulations through clustering.  

We performed K-means clustering on the exact mass parameters of the 400 binaries.  For clustering, we used the mass ratio $q \equiv m_2/m_1$ \footnote{From here on, $m_2$ is the smaller companion mass; in \autoref{fig:binaries}, it was the mass of the remnant of the secondary star -- the star which initially had a lower mass, but could end up as a more massive compact remnant at the end of binary evolution.} and the chirp mass $M_c \equiv (m_1 + m_2) \eta^{3/5}$, where $\eta$ is the symmetric mass ratio $\eta \equiv q (1+q)^{-2}$.  The chirp mass is chosen because it determines the gravitational-wave frequency evolution at the lowest order and is therefore the best-measured mass combination (see below).  For clustering purposes, we use a logarithmic coordinate on the chirp mass, $\ln (M_c / M_\odot - 0.8)$.  Simple k-means clustering \citep{MacQueen:1967}, which assigns each observation to a cluster with the closest mean, proves adequate for perfect classification on the true source parameters: every binary in \autoref{fig:truecluster} is correctly assigned to the right cluster.

\section{Measurement uncertainty}

In practice, inference on gravitational-wave signals permits only a limited accuracy of parameter estimation.  These limitations are due to significant correlations in the occasionally multi-modal parameter space of 15 or more parameters, including component masses and spins, as well as the binary's sky location and orientation.   Approximate techniques pioneered more than 20 years ago have demonstrated that the chirp mass is a relatively well-measured parameter for systems with a total mass of a few tens of solar masses or less, but other mass combinations, such as the mass ratio, can only be relatively poorly constrained \citep{CutlerFlanagan:1994,PoissonWill:1995}.  More recently, Bayesian techniques have been used to directly measure the posterior \acp{PDF} of the signal parameters given the observed noisy data \citep{S6PE,GW150914:PE}.  These techniques, encoded in the \texttt{LALINFERENCE} parameter-estimation pipeline \citep{Veitch:2014}, have been used to constrain the accuracy of parameter estimation on NS-NS, NS-BH, and BH-BH binaries in a variety of realistic contexts \citep[e.g.,][]{Vitale:2014,Veitch:2015,Littenberg:2015,Mandel:2015,Haster:2015IMRI,Farr:2016}.  

Here, we use these earlier results to generate mock posterior \acp{PDF} marginalised over all parameters other than $m_1$ and $m_2$.   We generate posterior samples in $\{$chirp mass $M_c$, symmetric mass ratio $\eta\}$ parameter space, given true values $(M_c^T, \eta^T)$, as follows:
\begin{eqnarray}
\vec{M_c} = M_c^T \left[1 + \alpha \frac{12}{\rho} (r_0  + \vec{r}) \right] \, ; \nonumber \\
\vec{\eta} = \eta^T \left[1 + 0.03 \frac{12}{\rho} (r_0' + \vec{r}') \right]\, .
\end{eqnarray}
Here, $r_0$ and $r_0'$ are random numbers drawn from the standard normal distribution and the corresponding terms encapsulate the shift in the mean of the posterior relative to the true value, while $\vec{r}$ and $\vec{r}'$ are independent and identically distributed arrays of such random numbers and represent the spread of the posterior.  The measurement uncertainty scales inversely with the signal-to-noise ratio $\rho$, which is drawn from the distribution $p(\rho) \propto \rho^{-4}$, which holds for isotropically distributed sources in a static universe, subject to the threshold $\rho \geq 8$ for detection.  The scaling $\alpha$ is motivated by analyses of mock data with the \texttt{LALINFERENCE} pipeline \citep[e.g.,][]{Littenberg:2015, Mandel:2015} and includes the impact of correlation with parameters describing arbitrary remnant spins; $\alpha=0.01$, 0.03, and 0.1 when $\eta^T \geq 0.1$, $0.1 > \eta^T \geq 0.05$, and $0.05 > \eta^T$, respectively.  Only posterior samples with $0.25 \geq \eta \geq 0.01$ are kept; no a priori cuts on individual masses are assumed, making this an intentionally somewhat conservative estimate of measurement uncertainty.

\begin{figure}
\centering
\includegraphics[width=0.45\textwidth]{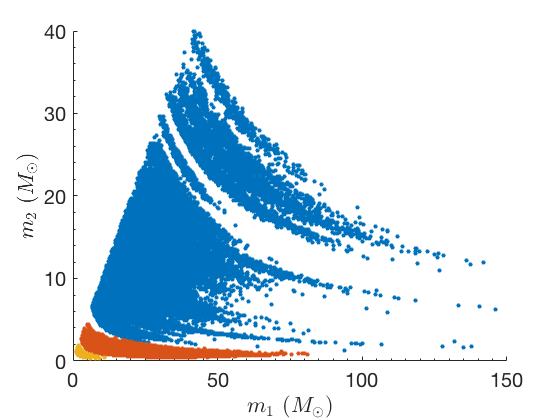}
\caption{500 posterior samples from each of 400 binaries in the catalog are placed on the same plot to demonstrate the impact of measurement uncertainty; samples are coloured based on the binary type of the source they are associated with, as in \autoref{fig:binaries}.}
\label{fig:uncertain}
\end{figure}

For each of the 400 mock binaries in our catalog shown in \autoref{fig:truecluster}, we generate between 500 and 2000 posterior samples in $\{m_1,m_2\}$ space, consistent with the typical posterior \acp{PDF} produced by \texttt{LALINFERENCE} \citep[e.g.,][]{S6PE}. In \autoref{fig:uncertain}, we overplot 500 posterior samples from each of the 400 measured events.  Each posterior distribution exhibits a typical `banana'-like shape, following contours of roughly constant chirp mass but spanning a range of values of the symmetric mass ratio.  The actual size of the posterior depends on the parameter values of the event and its simulated signal-to-noise ratio, while the distribution is randomly shifted relative to the true value so that the true value has a uniform probability of falling into every quantile of the posterior.  The combined posterior distributions appear to show an absence of a gap in between NS-NS and NS-BH binaries, and some overlap between NS-BH and BH-BH binaries; meanwhile, due to the lower merger density of higher-mass BH-BH binaries in our model, gaps appear at higher masses in the $\{m_1,m_2\}$ distribution.  

\begin{figure}
\centering
\includegraphics[width=0.45\textwidth]{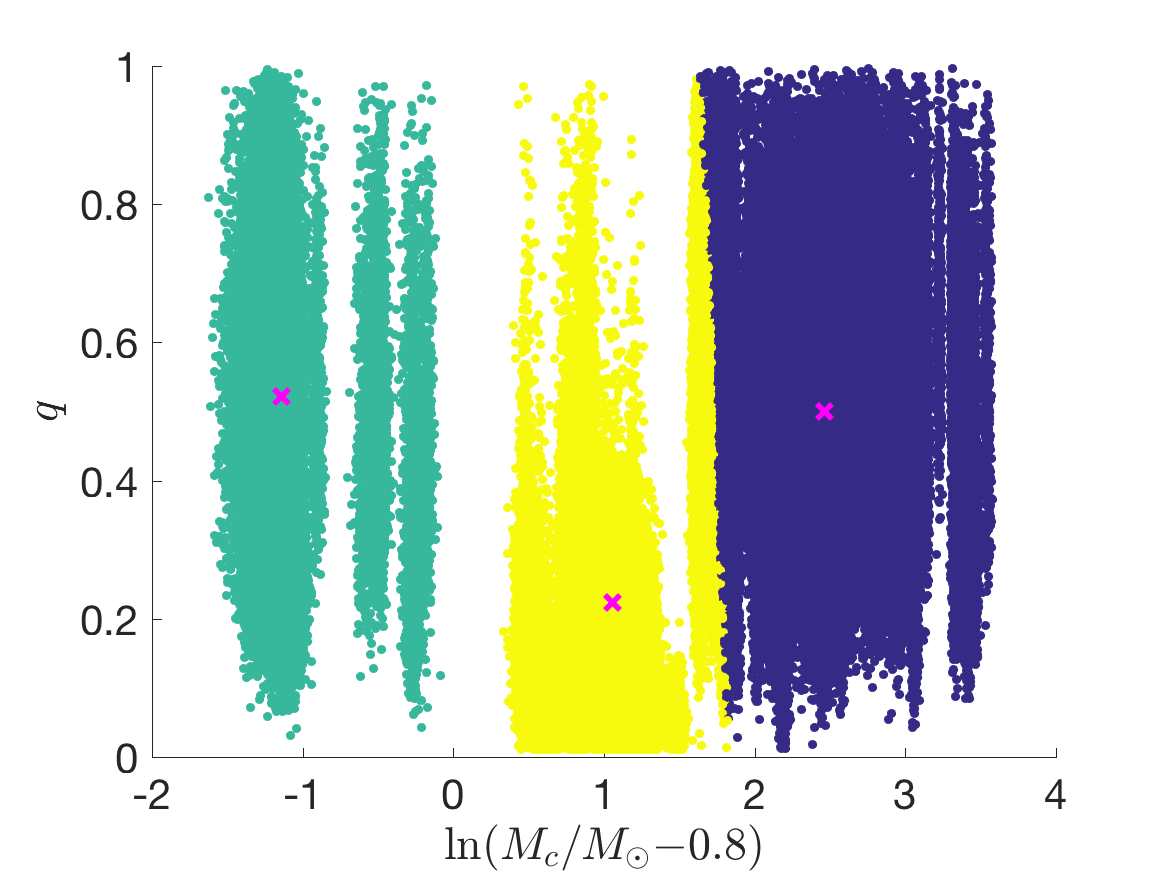}
\caption{K-means clustering on a ``bag'' of posterior samples (500 samples from each of 400 binaries).  Some misclassification is evident, and the cluster means, denoted with magenta Xs, no longer correspond to the true subpopulation clusters (cf.~\autoref{fig:truecluster}).}
\label{fig:uncertaincluster}
\end{figure}

As an indication of the difficulty of clustering on the observed population suffering from measurement uncertainties, we can apply the k-means clustering procedure described in the previous section to the full bag of $400 \times 500$ posterior samples plotted in \autoref{fig:uncertain}.  The result of this attempt is shown in \autoref{fig:uncertaincluster}.  The very large extent of the posteriors in the $q$ direction makes it difficult to estimate the true locations of the clusters, as evidenced by the shifting of the cluster means in mass ratio relative to their values in \autoref{fig:truecluster}.  It is still possible to cluster on the chirp mass, however, since it is relatively accurately measured.  

\begin{figure}
\centering
\includegraphics[width=0.45\textwidth]{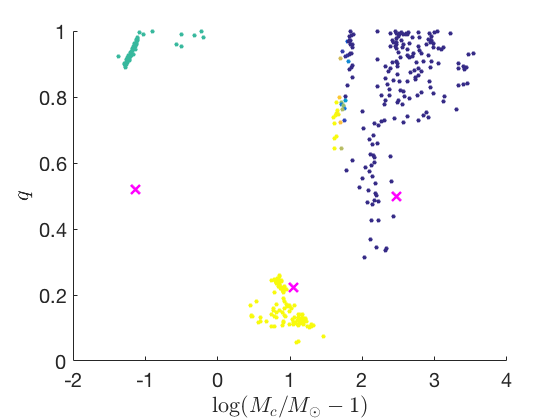}
\caption{K-means clustering on the same bag of posterior samples as in \autoref{fig:uncertaincluster}.  All 400 observations are displayed at their true parameter values, but the color reflects the mean cluster association of all posterior samples corresponding to each observation.}
\label{fig:uncertainclusterave}
\end{figure} 

However, despite the use of the logarithmic chirp mass coordinate $\ln (M_c / M_\odot - 0.8)$ to aid clustering, some of the BH-BH binaries are mis-identified by being associated with the NS-BH cluster.  This is most easily seen in \autoref{fig:uncertainclusterave}, in which we plot each observation at its true mass parameters, but colour it in based on the average cluster association of all corresponding posterior samples.  This figure shows yellow (classified as NS-BH) samples in the top-right ``BH-BH'' cluster (see also  
the right-most yellow strip of \autoref{fig:uncertaincluster}).  Consequently, the fractions of systems in the clusters approximately associated with NS-NS, NS-BH and BH-BH populations are $23\%$, $30\%$, and $47\%$, respectively.  The mis-classification errors now exceed those expected from Poisson (trinomial) statistics for 400 objects, and the classification becomes increasingly poor as the number of observations is reduced.

Of course, the approach described above is flawed because it fails to take advantage of all available information.  We lose key information by putting all posterior samples into a single ``bag'' and ignoring which observation each sample corresponds to.   For example, this means that we do not make use of the insight that some posteriors are very broad (and therefore not very useful for clustering), while others correspond to very precise measurements.  The right approach must build a hierarchical model out of the full observed population \citep[e.g.,][]{Hogg:2010,Bovy:2011,Mandel:2010stat,Farr:2013}, accounting for the individual measurement uncertainties, and search for subpopulations in this reconstructed population.  One possible implementation is described in the following section.

\section{Distribution inference and clustering}

We separate the problem into two parts: hierarchical modelling of the mass distribution based on a finite number of limited-accuracy observations, and clustering based on the inferred mass distribution.  There are many possible ways to parametrise the mass distribution model.  We follow \citet{ForemanMackey:2014} and \citet{massdistribution:2016} in choosing a piecewise-constant two dimensional distribution, i.e., we divide the mass space into rectangular bins and model the fraction of systems $n_k$ within each bin $k \in [1,K_\textrm{bins}]$.  In this case, we bin in $\ln m_1\, \times\, \ln m_2$ space, with square bins in log space.  We cover the range of component masses from $m=1 M_\odot$ to $m=181 M_\odot$ with a total of $K_\textrm{bins} =15 \times 15 = 225$ bins.

When $N$ independent observations are available, each represented by a data set $d^{(i)}$, the posterior probability density function on the distribution across the bins $\vec{n}\equiv\{n_k\}$ is given by \citep{Mandel:2010stat}
\begin{equation}
p(\vec{n}) \propto \pi(\vec{n}) \prod_{i=1}^{N} p(d^{(i)}|\vec{n})\, .
\end{equation}
Here
\begin{equation}
p(d^{(i)}|\vec{n}) = \int p(d^{(i)}|m_1^{(i)}, m_2^{(i)}) p(m_1^{(i)}, m_2^{(i)}|\vec{n}) dm_1^{(i)} dm_2^{(i)},
\end{equation}
where $p(d^{(i)}|m_1^{(i)}, m_2^{(i)})$ is the likelihood of observing the data $d^{(i)}$ given the specified masses \citep{Veitch:2014} and $p(m_1^{(i)}, m_2^{(i)}|\vec{n}) = n(m_1^{(i)}, m_2^{(i)})$ is the number density $n_k$ for the appropriate bin $k$ into which these masses fall.  In practice, we can replace the preceding integral over the likelihood function by a sum over the available posterior samples,  appropriately re-weighted by the prior used for individual event analysis \citep[see][for details]{Mandel:2010stat,Mandel:2016select}. 
Finally, $\pi(\vec{n})$ is the prior probability distribution on the fractions within the $K_\textrm{bins}$ bins.  Our prior is a stationary Gaussian process with a squared-exponential kernel, described in detail in \citet{massdistribution:2016}. This prior provides a crucial regularisation, favouring a smooth distribution when the data are sparse\footnote{If sharp edges are expected in the distribution, it would be preferable to use an alternative prior choice that does not disfavour such features.}, but allows the posterior to converge to the expected frequentist multinomial distribution when $N$ is large and the measurements are precise.  

\begin{figure*}
\centering
\includegraphics[width=0.3\textwidth]{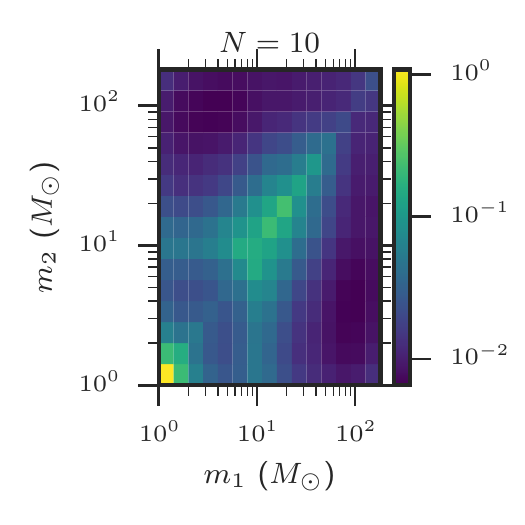}
\includegraphics[width=0.3\textwidth]{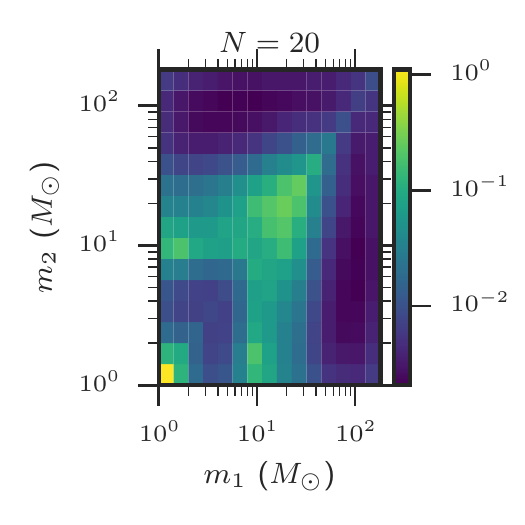}
\includegraphics[width=0.3\textwidth]{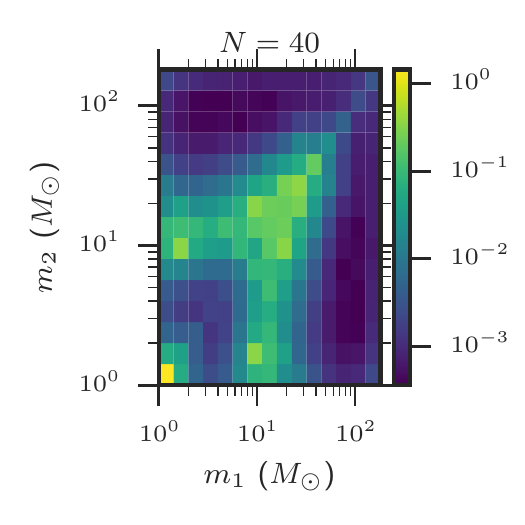}\\ 
\includegraphics[width=0.3\textwidth]{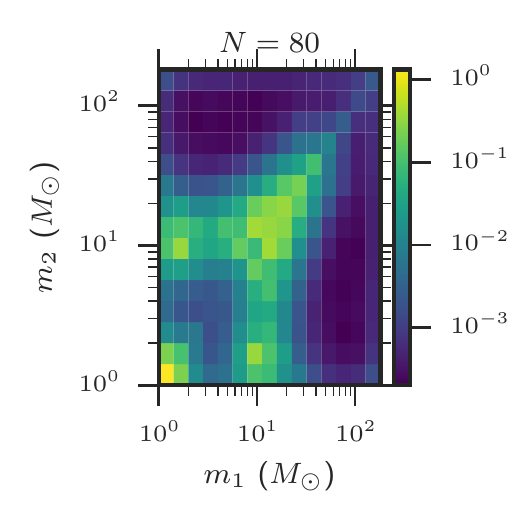}
\includegraphics[width=0.3\textwidth]{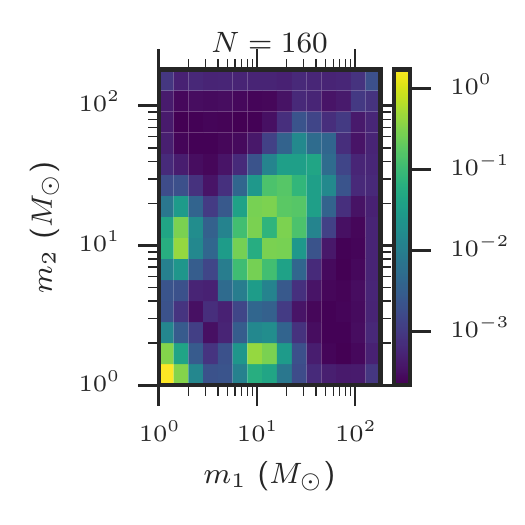}
\includegraphics[width=0.3\textwidth]{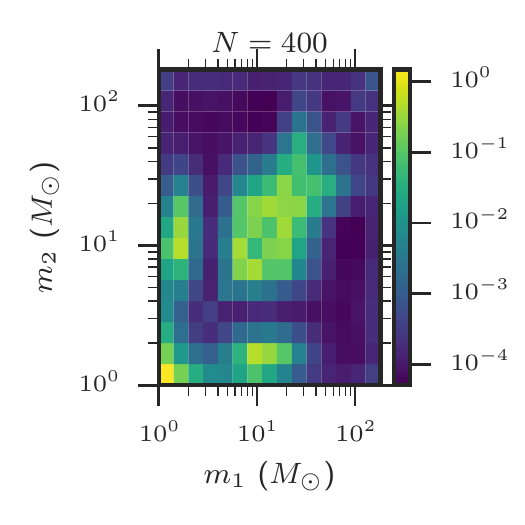}
\caption{Mean density inferred across mass space from mock observations using a binned distribution model with a Gaussian process prior for $N=10,\, 20,\, 40$ (top row, left to right) and  $80,\, 160,\, 400$ (bottom row, left to right) observations.}
\label{fig:GPmeans}
\end{figure*}

We compute the inferred distribution according to this hierarchical model from our mock data.  \autoref{fig:GPmeans} shows the posterior mean of the population density in each of the mass bins inferred from the full set of $400$ observations, as well as from smaller randomly drawn subsets to illustrate the gradual evolution of the accuracy of the inferred posterior.  Distinct NS-NS, NS-BH, and BH-BH clusters clearly appear around $40$ -- $80$ observations, consistently with the estimated requirement of $\sim 60$ observations made by \citet{Mandel:2015}. 

\begin{figure}
\centering
\includegraphics[width=0.45\textwidth]{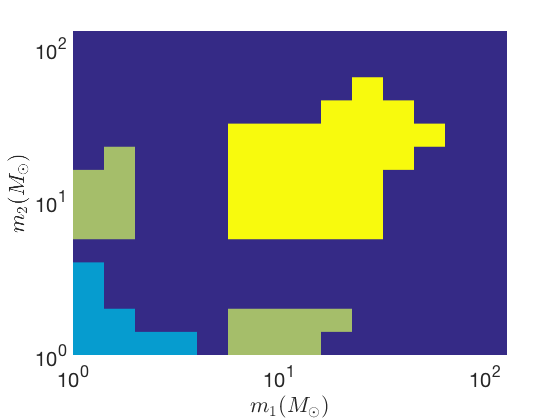}
\caption{Water-filling clustering on the mean estimates of the population fraction in each bin, as inferred from 400 mock observations.}
\label{fig:waterfilling}
\end{figure}

In order to identify specific clusters, we use a water-filling algorithm on the mean estimates 
of the population density in each bin \citep[see, e.g.,][for other proposed approaches to distributional clustering]{NielsenNock:2008, Van:2010, Applegate:2011}.  We gradually flood the posterior landscape until only three clusters stand above the water level over the $m_1 \geq m_2$ half of the plane.  Clusters here are defined as sets of bins such that all elements of a cluster are connected through shared edges, but such connections do not exist between distinct clusters.  Some of the posterior ends up in the under-water bins; the clustering is deemed successful only when under-water bins account for no more than a few percent of the posterior.  This happens starting with $N = 80$ for the plots in \autoref{fig:GPmeans}.  As an example, \autoref{fig:waterfilling} shows the results of applying the water-filling clustering strategy to the distribution inferred from $N = 400$ observations (mirrored across $m_1=m_2$ for plotting).  In this case, the NS-NS, NS-BH, and BH-BH subpopulations contain $23\%$, $25\%$, and $51\%$ of the population, respectively, while less than $2\%$ of the posterior is under-water.

In general, the appropriate number of clusters does not need to be assumed in advance, but should be chosen from the data during the water-filling stage.   Specifically, the amount of water used for flooding can be optimised against the flooded area.  Flooding should continue only while the flooded area grows rapidly with a modest increase in the posterior volume (the amount of water used for flooding), with the remaining above-water areas identified as clusters.

\begin{figure*}
\centering
\includegraphics[width=0.45\textwidth]{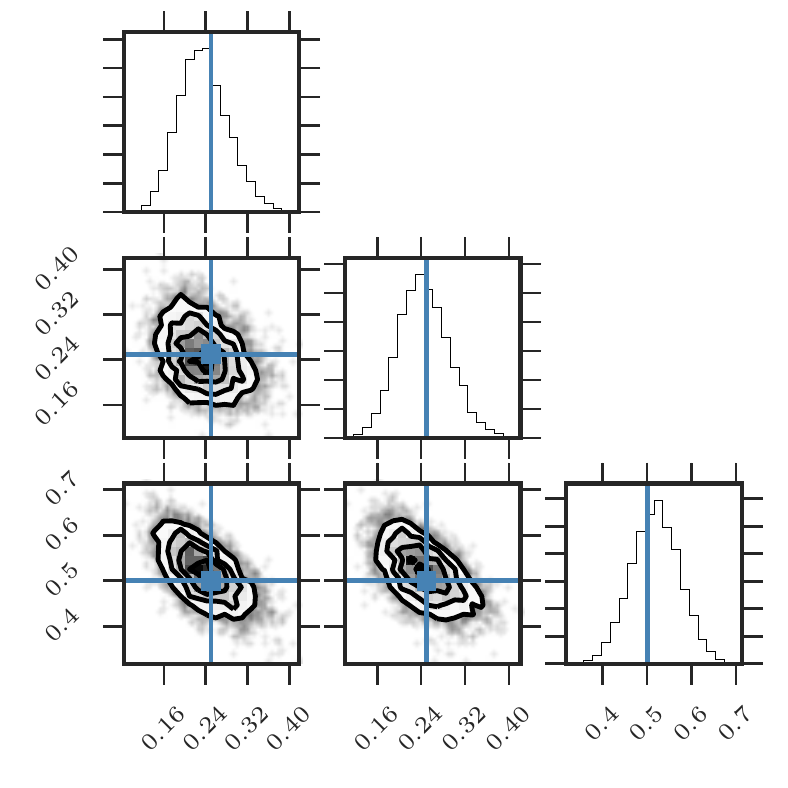}
\includegraphics[width=0.45\textwidth]{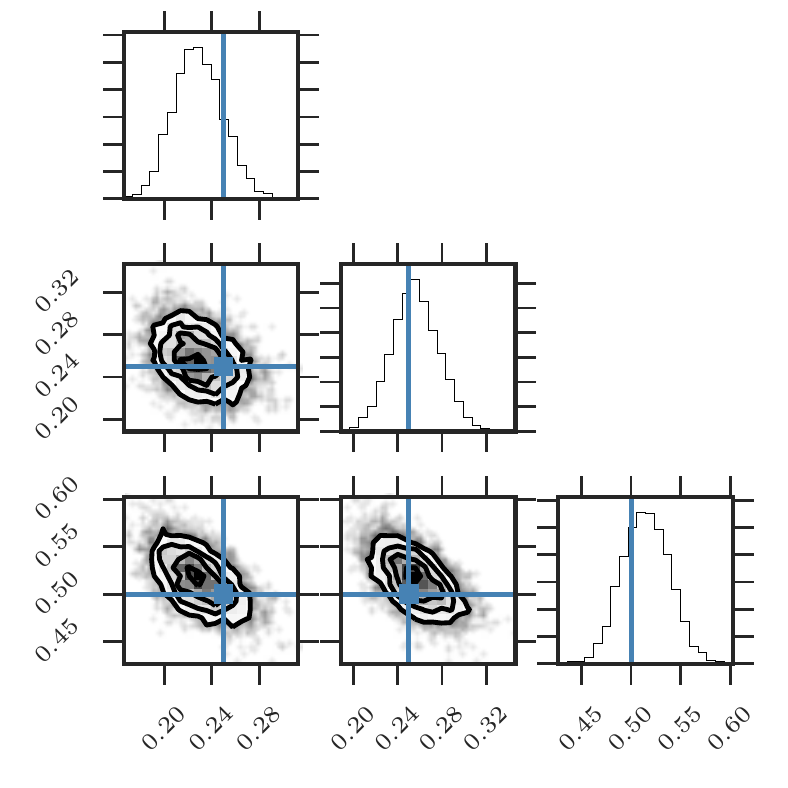}
\caption{Triangle plots for the posteriors on the inferred fraction of events in each of the NS-NS, NS-BH, and BH-BH subpopulations (ordering from left to right, and top to bottom); blue lines denote the fractions used to randomly draw the events being clustered.  (Left) The posterior after 80 mock observations.  (Right) The posterior after 400 mock observations.}
\label{fig:trianglepost}
\end{figure*}

We can obtain estimates of the statistical uncertainty on the inferred posterior fraction in each cluster by taking advantage of the full \acp{PDF} on the fractional mass distribution within each bin.  We use the cluster boundaries provided by the water-filling clustering algorithm and compute the posterior on the total mass density within each cluster identified with the NS-NS, NS-BH, and BH-BH subpopulations.  To be precise, given the posterior over all $\vec{n}$, we simply add the individual posteriors on the sums of those $n_k$ which fall into a particular cluster; we do not account for the uncertainty in the cluster boundaries when computing these cluster fraction posteriors.  The triangle plots for the cluster fraction posteriors are shown in \autoref{fig:trianglepost}.\footnote{For this figure we associated each under-water bin with a neighbouring cluster; this does not impact the results other than ensuring that the three fractions sum to $1$.}    For both 80 (left) and 400 (right) observations, the uncertainty in the inferred fraction of each subpopulation is within the expected fluctuation in random-draw statistics from a trinomial distribution, as predicted by \citet{Mandel:2015}.

\vspace{0.2in}

We have presented a practical technique for clustering observations suffering from significant measurement uncertainty.  We demonstrated its functionality on the mass parameter space and showed that a realistic population of merging compact binaries could be accurately clustered into NS-NS, NS-BH, and BH-BH subpopulations.  The number of observations required for accurate clustering will depend on how well-separated the true subpopulations are, on the actual fractions of events in each subpopulation and on the size of measurement uncertainties.  Our example indicates that $\sim 20$ observations per subpopulation are more than sufficient for accurate clustering on the modelled population.  We have confirmed that this number of observations per subpopulation is sufficient for accurate clustering even when the ratio between the numbers of events in different subpopulations is more extreme, e.g., $1:5:50$ rather than $1:1:2$.  With sufficient observations, it should be possible to use this technique to cluster on any population with multiple modes separated by lower-density regions in parameter space (gaps), even if the measurement uncertainty on individual observations is larger than the width of the gaps.  It is straightforward, though computationally expensive, to extend this technique to higher-dimensional analyses, e.g., to include spin information along with mass information, which could help to distinguish isolated and dynamical formation channels for binary black holes \citep{GW150914:astro,BBH:O1}.

\section*{Acknowledgments}
  
IM and WF acknowledge partial financial support from STFC.

\bibliographystyle{mnras}
\bibliography{Mandel}
\label{lastpage}
\end{document}